\documentclass[conference]{IEEEtran}

\usepackage{cite}
\usepackage{amsmath,amssymb,amsthm}
\usepackage{url}

\usepackage{graphicx}
\usepackage{xcolor}
\usepackage{booktabs}
\usepackage{algorithm}
\usepackage{algpseudocode}

\IEEEoverridecommandlockouts

\begin{document}

\title{Interference-Aware Tone Selection for 
Bluetooth Channel Sounding}

\author
{Taewoo Kim\textsuperscript{\textdagger} and Vishrant Tripathi\textsuperscript{\textdagger}
\thanks{
\textsuperscript{\textdagger} Electrical and Computer Engineering, Purdue University, West Lafayette, USA, Email: kim5354@purdue.edu, tripathv@purdue.edu}
}


\maketitle
\begin{abstract}
Bluetooth Channel Sounding enables distance estimation using phase measurements across multiple frequencies. However, coexistence with Wi-Fi and other 2.4 GHz radios can fragment the available frequency band, making reliable ranging difficult when only a limited number of tones can be selected. Drawing on sparse antenna-array theory, we show that interference avoidance alone is insufficient: the spacing of selected tones can introduce repeated peaks in the ranging response, leading to estimation errors. We propose a lightweight, interference-aware tone-selection method that jointly considers frequency occupancy, tone spacing, and frequency span. Our method constructs a 15-tone map using a three-step rule without ranging measurements, ground-truth distance, or online candidate-map search. We evaluate the method using simulations of 900 indoor environments, a public Bluetooth Channel Sounding dataset, and measurements from commercial Nordic nRF54L15 hardware. Compared with random interference avoiding baseline, our proposed method improves ranging success percentage by 39 points under mixed interference, 29 points on the public dataset, and 82 points in hardware experiments. These results demonstrate the importance of tone selection for practical Bluetooth Channel Sounding.
\end{abstract}


\vspace{-2mm}
\section{Introduction}
\label{sec:intro}
\vspace{-2mm}
Accurate distance estimation between wireless devices is important for applications such as digital car keys, indoor navigation, industrial automation, asset tracking and item finding. Bluetooth has traditionally estimated distance using received signal strength (RSSI), which provides a coarse and unreliable measure of proximity, with accuracy around 3-5m \cite{faragher}. During ranging, received power depends not only on distance, but also on fading, obstruction, multipath propagation, and antenna orientation, making it difficult to distinguish devices that are physically close from those that are farther away. RSSI-based ranging is also vulnerable to relay attacks, in which an adversary amplifies a relayed signal to make a distant device appear nearby \cite{gentner,francillon,staat}. To address these limitations, Bluetooth Channel Sounding (CS) was introduced in version 6.0 of the Bluetooth Core Specification \cite{btcore60}. Bluetooth CS estimates distance using phase measurements across multiple frequencies and round-trip timing, enabling more accurate ranging, around 1m \cite{sigcs}, while also incorporating mechanisms to mitigate relay attacks. By supporting fine-grained distance estimation within the existing Bluetooth ecosystem, CS has the potential to enable a broader range of proximity-based applications without requiring a separate radio technology.

Since Bluetooth CS recovers distance from phase measurements of unmodulated tones transmitted across multiple frequencies in the 2.4~GHz band, the choice of frequencies used directly affects the information available for distance estimation, particularly in multipath environments. 
The Bluetooth specification defines 79 frequency indices, each corresponding to a sounding tone, of which 72 are available for sounding. An implementation can select between 15 and 72 tones for each ranging estimate~\cite{btcore60}.
However, the specification does not prescribe how these tones should be selected, and reference implementations from silicon vendors typically use all 72 \cite{nordiccs, silabs}. Selecting fewer tones can reduce sounding overhead and radio energy consumption, which is particularly important for battery-constrained devices. Careful tone selection can also improve coexistence with Wi-Fi and other technologies sharing the 2.4~GHz band by avoiding frequencies heavily affected by interference. These benefits, however, depend on selecting tones that preserve ranging accuracy while limiting interference. This raises a fundamental question: \textbf{how should a limited number of tones be selected to achieve accurate and robust ranging across diverse environments?}

Prior studies have shown that coexistence between Bluetooth and Wi-Fi can substantially degrade WLAN performance, with packet loss reaching $45\%$ in the presence of a single Bluetooth piconet and 57-$84\%$ with multiple piconets \cite{golmie}. 
Similarly, Wi-Fi interference degrades the Bluetooth ranging: with all 72 tones and Wi-Fi channels 1, 6 and 11 all occupied, we observe the median error in our simulations rising from 0.52~m to 2.33~m, at a realized signal-to-interference ratio of $-5$~dB.
\color{black}These results motivate accounting for coexistence when selecting sounding tones.

In our evaluation, we select the minimum of 15 tones permitted by the specification \cite{btcore60}, reducing sounding overhead while seeking to avoid frequencies occupied by Wi-Fi. However, selecting the best subset is challenging: even among the 72 available tones, there are $\binom{72}{15}$ possible subsets of 15 tones. Exhaustively evaluating these choices is impractical, and the best subset must adapt to different physical sensing conditions and balance interference avoidance with the frequency coverage needed for accurate ranging. Moreover, interference can occupy different portions of the band across deployments and regulatory regions, making it difficult to rely on fixed tone selection. We make three key contributions in this work.

\begin{itemize}
\setlength{\itemsep}{0pt}
\setlength{\parskip}{0pt}
\item \textbf{Fundamental Limitations.} Selecting tones to avoid interference can leave large gaps between frequencies, making it difficult to distinguish the true device distance from incorrect estimates. 
We provide a simple spacing condition for uniformly spaced tones and explain how irregular tone placement can introduce secondary peaks in the distance response.

\item \textbf{Practical Tone Selection.} We develop a practical tone-selection method that balances interference avoidance with frequency coverage, and evaluate it using a generalized interference simulation. We then precompute tone selections for different interference-channel configurations in a lookup table, enabling devices to configure their sounding tones at run time \textit{without online optimization}.

\item \textbf{Simulations and Experimental Validation.} We evaluate our proposed method across 900 simulated rooms using a commercial distance-estimation stack. We further assess its performance using a public dataset and measurements collected with commercial hardware, comparing its ranging accuracy against alternative tone-selection strategies.
\end{itemize}

\vspace{-2mm}
\section{Related Work}
\label{sec:related}
\vspace{-1mm}
Phase-based ranging (PBR) over Bluetooth has been studied for accurate distance estimation \cite{zand}, including estimation with missing or interfered tones \cite{sheikh}. Strong Wi-Fi interference has been shown to increase random ranging error by approximately fourfold and degrade PBR performance beyond 10~m \cite{kravets}. These studies focus on ranging estimation and interference characterization, whereas our work addresses sounding-tone selection under frequency occupancy constraints.

Tone selection is closely related to sparse-array design and frequency placement. Prior work investigates sidelobes in randomly spaced arrays \cite{lo,steinberg}, minimum-redundancy arrays \cite{moffet}, and the optimization of thinned arrays \cite{haupt}. Similar principles apply to stepped-frequency radar \cite{levanon,axelsson} and multicarrier phase ranging \cite{liu}, where frequency spacing affects range ambiguity. Building on these frequency-placement principles, our method constructs Bluetooth CS tone maps based on interference while accounting for frequency span and tone spacing without iterative optimization.

Adaptive frequency hopping \cite{poirot} and combinatorial bandit approaches \cite{chen13,chen14} select channels using feedback from prior channel observations. Such approaches require observations to guide channel selection and may need to adapt when interference conditions change. In contrast, our method constructs tone maps directly from the occupancy configuration without ranging feedback, ground-truth distance, or online candidate-map evaluation. For typical interference patterns, these maps can be precomputed and stored in a lookup table.

\vspace{-2mm}
\section{Channel Sounding}
\label{sec:problem}
\vspace{-1mm}
\begin{figure}[t]
\centering
\includegraphics[width=0.88\columnwidth]{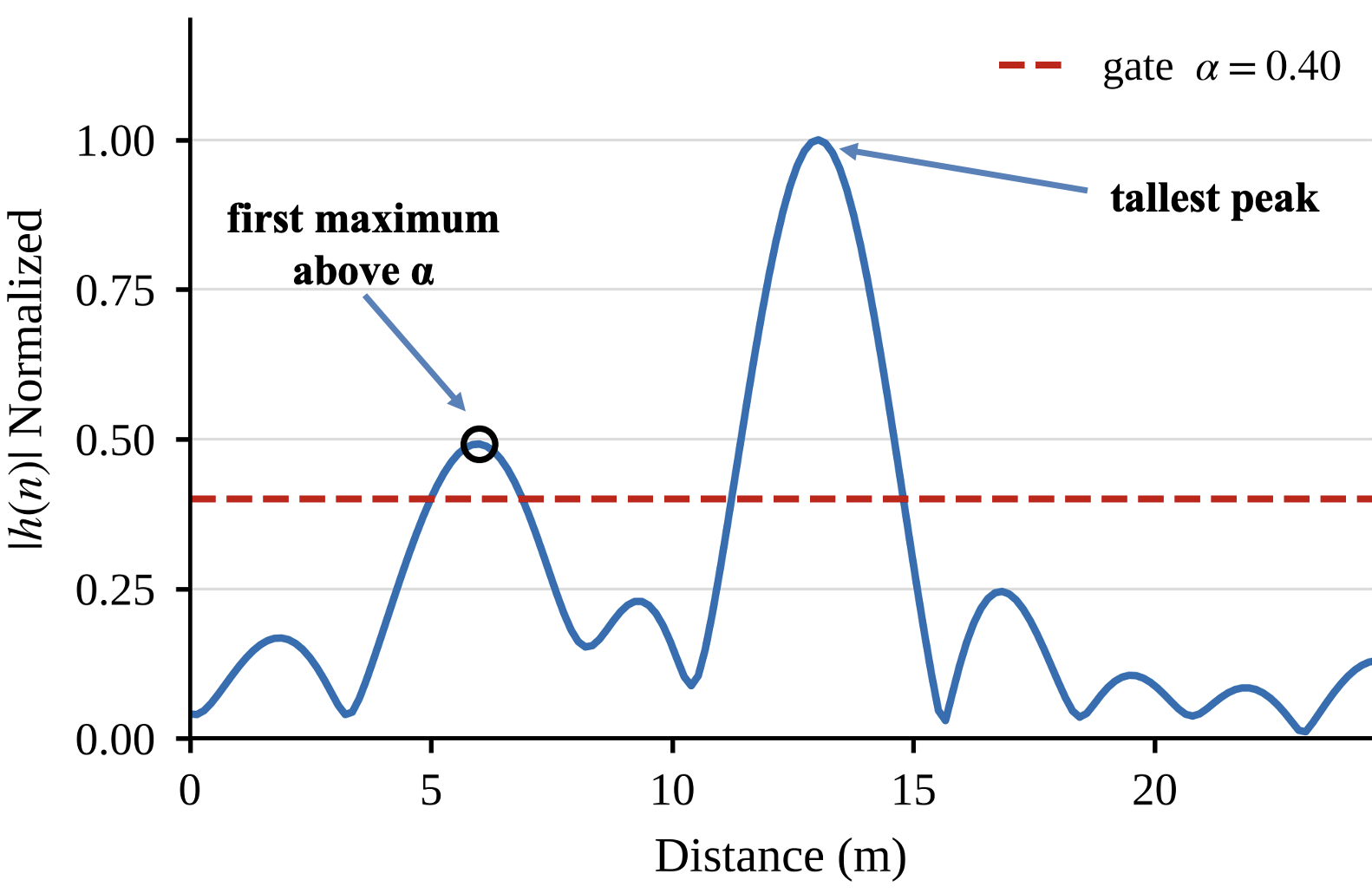}
\caption{Leading-edge distance estimation from a normalized distance response. The dashed line indicates the detection threshold at $\alpha=0.40$ of the global maximum. The estimator selects the first local maximum above this threshold rather than the strongest peak.}
\label{fig:highlevel}
\end{figure}

\subsection{Signal Model and Estimator}

In a Bluetooth CS procedure, the initiator and reflector exchange an unmodulated continuous-wave tone at each selected frequency. Frequency index $k$ corresponds to $f_k = f_0 + k\,\Delta f$, where $f_0 = 2402$~MHz and $\Delta f = 1$~MHz. Each device records the in-phase and quadrature components of the received tone. Combining these measurements cancels the local oscillator offset, leaving the round-trip phase. For a propagation path of length $d$, the round-trip delay is $\tau = 2d/c$, which introduces a phase rotation of $-2\pi f_k \tau$ at frequency $f_k$. With $Z$ propagation paths in the room, the measurements at the selected tones are modeled as
\begin{equation}
H(k) = \sum_{q=0}^{Z-1} a_q e^{-j2\pi f_k \tau_q} + w_k,
\qquad k \in S,
\label{eq:response}
\end{equation}
where $a_q$ and $\tau_q$ denote the complex amplitude and round-trip delay of path $q$, respectively; $w_k$ captures measurement noise and interference from other radios at index $k$; and $S$ denotes the selected frequency-index map. \textit{We set $|S| = K = 15$ throughout this paper, although our approach is generalizable to arbitrarily sized tone sets.}


Let $\mathcal{P}$ denote the 72 permitted tones. We define the \emph{occupancy set} $W$ as the set of tones affected by significant interference and the \emph{free set} as $L_0=\mathcal{P}\setminus W$. 

As a baseline, random interference-avoiding tone (RIT) selection uniformly selects $K=15$ distinct tones from $L_0$. If fewer than $K$ free tones are available, RIT retains all free tones and randomly fills the remaining positions from $\mathcal{P}\setminus L_0$.

Our tone selection method, discussed in Section~\ref{sec:rule}, takes the occupancy set $W$ as the only input and returns a 15 tone frequency index map $S$ for channel sounding.

The receiver transforms the frequency-domain measurements in (\ref{eq:response}) into a distance response using an inverse discrete Fourier transform (IDFT). Because measurements are available only at the selected tones \(k \in S\), the receiver constructs an array of length \(N_{\mathrm{FFT}}\), placing each measurement \(H(k)\) at its corresponding frequency index and setting all other entries to zero. The resulting distance response is given by
\begin{equation}
h(n) = \sum_{k \in S} H(k) \exp\left( \frac{j 2 \pi k n}{N_{\mathrm{FFT}}} \right).
\label{eq:ifft}
\end{equation}

Each propagation path contributes a frequency-dependent phase term \(e^{-j2\pi f_k\tau_q}\) to \(H(k)\). The IDFT coherently combines these measurements, producing a peak near the sample corresponding to the path's round-trip delay. For frequency spacing \(\Delta f\), delay sample \(n\) corresponds to a round-trip delay \(\tau_n = n/(N_{\mathrm{FFT}}\Delta f)\) and a one-way distance \(d_n = cn/(2N_{\mathrm{FFT}}\Delta f)\).


The receiver estimates distance from the magnitude of the distance response \(|h(n)|\) (see Fig.~\ref{fig:highlevel}). The earliest propagation path corresponds to the shortest propagation distance and therefore determines the true arrival. Ideally, this path produces the first detectable peak in the distance response. In practice, however, the profile may contain other local maxima caused by multipath propagation, measurement noise, interference, and the limited set of sampled frequencies. These maxima can occur before the true arrival and may be mistaken for an earlier propagation path. Fig.~\ref{fig:highlevel} illustrates the leading-edge estimator using a normalized distance response with multiple local maxima. The dashed line indicates the detection threshold, and the estimator selects the first local maximum exceeding this threshold rather than the strongest peak. Consequently, a sidelobe appearing before the true-arrival peak can produce an underestimated distance if it exceeds the detection threshold.

\vspace{-1mm}
\subsection{Frequency Fragmentation}
Interference can occupy different portions of the available band, leaving fragmented groups of usable tones in ${L}_0$. Consequently, the selected tones may be irregularly spaced, and their placement affects the distance response in (\ref{eq:ifft}). This effect is closely related to the sidelobes and grating lobes studied in sparse antenna arrays \cite{lo,steinberg,moffet,haupt}.


\begin{figure}[t]
\centering
\includegraphics[width=0.88\columnwidth]{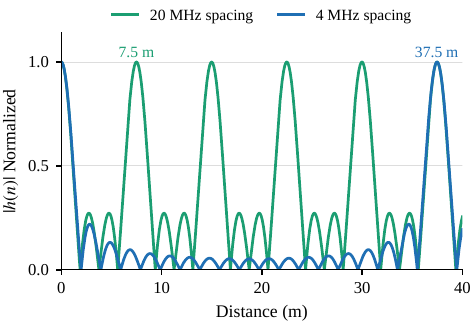}
\caption{Normalized distance responses for uniformly spaced tones with spacings of 20 and 4~MHz and a single arrival at 0~m. The responses exhibit repeated peaks every 7.5 and 37.5~m, respectively, illustrating how larger tone spacing reduces the unambiguous ranging distance.}
\label{fig:repeat}
\end{figure}

Using results from this line of work, we can look at how tone placement produces repeated peaks in the distance response. When the selected tones are equally spaced by $g$ MHz, increasing the one-way distance by
\begin{equation}
d_{\mathrm{alias}}=\frac{c}{2g\times10^6}
=\frac{149.896}{g}\ \mathrm{m}
\label{eq:alias_distance}
\end{equation}
adds a phase change of $2\pi$ between adjacent tones. The relative phase progression across the selected tones is therefore unchanged, producing a repeated peak in the distance response. Fig.~\ref{fig:repeat} compares normalized distance responses for uniformly spaced tones with 20 and 4~MHz spacing, assuming a single arrival at 0~m. The responses exhibit repeated peaks every 7.5 and 37.5~m, respectively, consistent with (3). In this idealized single-path example, the repeated peaks have the same magnitude as the original peak. A larger tone spacing therefore places ambiguous peaks closer to the true arrival, increasing the risk that the leading-edge estimator selects an incorrect peak within the target ranging distance.


To avoid this periodic ambiguity within a target ranging distance $d_{\max}$, the alias distance must satisfy $d_{\mathrm{alias}}>d_{\max}$. Equation~ (\ref{eq:alias_distance}) then requires
$
g < \frac{149.896}{d_{\max}}.$
For integer-MHz spacing, the largest spacing satisfying this strict inequality is
\begin{equation}
g^\star=
\left\lceil\frac{149.896}{d_{\max}}\right\rceil-1.
\label{eq:gstar}
\end{equation}
For example, when $d_{\max}=16$~m, $g^\star=9$~MHz. Equally spaced tones with gaps of at most 9~MHz place the first repeated peak beyond the target range (16~m), whereas larger gaps than 9~MHz produce a repeated peak inside the target range. This bound characterizes periodic ambiguity for uniformly spaced tones.

In practice, however, interference and tone selection both produce irregular rather than uniformly spaced tone sets, and such sets do not generally produce equally tall, periodically repeated peaks. Nevertheless, their distance responses can contain secondary peaks whose locations and magnitudes depend on the full pattern of selected tones. A secondary peak can still cause an underestimated distance if it occurs before the true-arrival peak and exceeds the estimator's detection threshold. Thus, avoiding the periodic ambiguity of uniformly spaced tones does not by itself eliminate false detections for irregular tone sets. The tone-selection method in Section~\ref{sec:rule} addresses this constraint while accounting for the available tones.



\section{The proposed method}
\label{sec:rule}


\begin{algorithm}[h]
\caption{Interference-aware tone selection}
\label{alg:rule}
\small
\begin{algorithmic}[1]
\Require Occupied tones $W$, permitted tones $\mathcal{P}$, tone count $K$, spacing bound $g^{\star}$, tolerance factor $p \in [0,1]$
\Ensure Selected tone map $S \subseteq \mathcal{P}$, $|S|=K$

\State $L_0 \gets \mathcal{P}\setminus W$
\State Partition $L_0$ into maximal groups $\mathcal{G}$ with consecutive gaps $\le g^{\star}$

\If{$\exists\, G \in \mathcal{G}$ with $|G|\ge K$}
    \State $\mathcal{A} \gets
    \arg\max_{G\in\mathcal{G},\,|G|\ge K}
    \bigl(\max G-\min G\bigr)$
    \State $\ell \gets \min\mathcal{A}$; $u \gets \max\mathcal{A}$
\Else
    \State $\mathcal{A}\gets\mathcal{P}$
    \If{$L_0\neq\emptyset$}
        \State $\ell\gets\min L_0$; $u\gets\max L_0$
    \Else
        \State $\ell\gets\min\mathcal{P}$; $u\gets\max\mathcal{P}$
    \EndIf
\EndIf

\State $\sigma\gets(u-\ell)/(K-1)$; $S\gets\emptyset$
\For{$t=0$ \textbf{to} $K-1$}
    \State $\tau\gets\ell+t\sigma$
    \State $\mathcal{C}\gets\mathcal{A}\setminus S$
    \State $\mathcal{C}_0\gets
    \{a\in\mathcal{C}\cap L_0:|a-\tau|\le p\sigma\}$
    \If{$\mathcal{C}_0\neq\emptyset$}
        \State $s\gets\arg\min_{a\in\mathcal{C}_0}|a-\tau|$
    \Else
        \State $s\gets\arg\min_{a\in\mathcal{C}}|a-\tau|$
    \EndIf
    \State $S\gets S\cup\{s\}$
\EndFor
\State \Return $S$
\end{algorithmic}
\end{algorithm}
We design a three-step rule to select $K$ sounding tones from the permitted tones. As summarized in Alg.~\ref{alg:rule}, the rule first identifies the free set $L_0$, then selects a frequency window subject to the spacing bound $g^\star$, and finally places the tones across that window. Its only runtime input is the occupancy set \(W\); it does not require ground-truth distance or online ranging optimization.

\subsection{Window Selection}



Alg.~\ref{alg:rule} partitions the free set ${L}_0$ into maximal groups separated by gaps greater than $g^\star$. Each group contains free tones whose adjacent spacings do not exceed $g^\star$. Among groups containing at least $K$ free tones, Alg.~\ref{alg:rule} selects the one with the widest frequency span. This choice preserves a wide frequency span for distance resolution while limiting large gaps between available tones, helping reduce the risk of repeated peaks. We denote the lowest and highest frequency indices in the selected window by $\ell$ and $u$, respectively.

\vspace{-1mm}
\subsection{Tone Placement}
\label{sec}

Within the selected window $[\ell,u]$, Alg.~\ref{alg:rule} first places $K$ equally spaced candidate positions. The spacing between these positions is $\sigma=(u-\ell)/(K-1)$. These positions serve as references for selecting the actual sounding tones, which may not be equally spaced.

The goal is to distribute tones across the selected frequency span while avoiding interference. We introduce a parameter \(p\in[0,1]\) that defines the search radius \(p\sigma\) around each candidate position, where \(\sigma\) is the spacing between adjacent candidate positions. For each candidate position, Alg.~\ref{alg:rule} selects the nearest unused free tone within this radius, if available. Otherwise, it selects the nearest unused candidate  outside.

A larger \(p\) value provides greater flexibility to avoid occupied tones but also permits greater deviations from the uniform tone spacing. Non-uniform frequency placement can suppress periodic range ambiguities associated with uniform spacing, but may introduce stronger sidelobes at incorrect candidate distances~\cite{Si2020,Wang2016}. For our experiment, we set $p=0.5$.

When no group contains at least $K$ free tones, Alg.~\ref{alg:rule} enters the \textit{fill branch}. Instead of restricting selection to a single group, it considers all permitted tones and distributes the candidate positions across the span of the remaining free tones, or the full permitted span if none remain. The same $p\sigma$ rule is then applied, allowing occupied tones when necessary to complete the $K$-tone map.
Although the \textit{fill branch} ensures that $K$ permitted tones are selected, it cannot guarantee interference-free selection. In particular, when fewer than $K$ free tones remain overall, selecting occupied tones is unavoidable.



\vspace{-2mm}
\section{Evaluation Setup}
\label{sec:eval}
\vspace{-1mm}
We evaluate our proposed method against RIT selection using simulated indoor rooms, a public measurement dataset, and measurements collected with commercial Bluetooth hardware. Both methods use $K=15$ tones from the 72 permitted tones under matched occupancy conditions.

\textbf{Indoor Environments:} We generate 900 rooms with different dimensions and uniformly sample Tx--Rx locations within these rooms. The generated rooms represent different typical geometries and have varying wall reflection coefficients between 0.15 and 0.98. Specular components are generated using the image-source method through third-order reflections [27]. Diffuse components follow an exponential power-delay profile consistent with the IEEE 802.11n and IEEE 802.15.4a channel models [28], [29]. True Tx--Rx separations are uniformly distributed between 3 and 16m. 

\textbf{Interference Sources: }
Each simulated environment contains up to three randomly selected Wi-Fi channels from channels 1--11, zero to six IEEE 802.15.4 networks with 2 MHz bandwidth [8], and zero to ten Bluetooth piconets with 1 MHz bandwidth. With probability 0.15, an additional 25 MHz-wide emitter near 2450 MHz represents a source such as a microwave oven or video sender. In addition to these random mixtures, we evaluate seven Wi-Fi occupancies formed from channels 1, 6, and 11. These cases represent common access-point configurations [11] and allow direct comparison with the hardware experiment in Section~\ref{sec:bench}. 

\textbf{SINR:} At each occupied index, we add an independent complex Gaussian interference sample. Its power is set so that the single-sounding signal-to-interference ratio is uniformly distributed between $-10$ and $10$ dB. The Bluetooth specification permits between 1 and 255 soundings per tone \cite{btcore60}. We use three soundings and average their measurements.



\textbf{Evaluation Metric:} We evaluate ranging accuracy using the fraction of estimates whose absolute error is smaller than one resolution cell. The nominal distance resolution for a bandwidth $B$ is $c/(2B)$. Using the full 74~MHz span of the permitted spectrum gives a resolution cell of 2.026~m. We use this threshold for all tone-selection methods to ensure a consistent comparison. We also report success rates at additional error thresholds of 1.0, 2.5, and 3.0~m in Fig.~\ref{fig:grid}, showing that our method outperforms RIT selection across all tested thresholds.


\vspace{-1mm}
\section{Results}
\label{sec:results}
\vspace{-1mm}
We first compare our proposed tone-selection method (Alg.~\ref{alg:rule}) with RIT selection across 60 interference mixtures. We then examine seven occupancy  configurations involving commonly used 2.4~GHz Wi-Fi channels 1, 6, and 11~\cite{farshad}. For these configurations, the 15-tone maps generated by Alg.~\ref{alg:rule} are precomputed and stored in a lookup table (Table~\ref{tab:maps}), enabling tone selection without online optimization.

\subsection{Interference Mixtures}

Table~\ref{tab:real} reports the share of measurements whose absolute ranging error falls within one resolution cell. Averaged across all interference configurations, our method achieves 64.7\%, compared with 25.2\% for RIT selection, a gain of 39.5 percentage points. The gain depends on the number of occupied tones. With fewer than 20 occupied tones, the proposed method achieves 68.1\%, compared with 34.1\% for RIT selection. When 20--39 tones are occupied, the proposed method achieves 74.5\%, compared with 24.9\%. At higher occupancy, the benefit is smaller: 58.2\% versus 22.5\% for 40--59 occupied tones, and 56.0\% versus 33.4\% for at least 60 occupied tones.

When fewer tones are occupied, our proposed method can select interference-free tones over a wide frequency span while avoiding large gaps. As occupancy increases, fewer free tones remain, forcing our method to use a narrower frequency span or select occupied tones through the \textit{fill branch}. This reduces the ranging improvement over RIT selection.

\begin{table}[t]
\caption{Ranging Success Rates On Interference Mixtures}
\label{tab:real}
\centering
\setlength{\tabcolsep}{5pt}
\footnotesize
\begin{tabular}{lccc}
\toprule
Occupied tones & RIT & Algorithm 1 & $\Delta$ \\
\midrule
$<20$       & 34.1\,\% & 68.1\,\% & $+34.0$ \\
$20$--$39$  & 24.9\,\% & 74.5\,\% & $+49.6$ \\
$40$--$59$  & 22.5\,\% & 58.2\,\% & $+35.7$ \\
$\ge 60$    & 33.4\,\% & 56.0\,\% & $+22.6$ \\
\midrule
All         & 25.2\,\% & 64.7\,\% & $+39.5$ \\
\bottomrule
\end{tabular}
\\[3pt]
\footnotesize Each configuration has multiple Wi-Fi transmitters, IEEE 802.15.4 networks, Bluetooth piconets and an occasional 25~MHz emitter. $\Delta$ denotes the improvement in percentage points.
\end{table}

\subsection{Wi-Fi Channel Occupancy}

\begin{figure}[ht]
\centering
\includegraphics[width=0.48\textwidth]{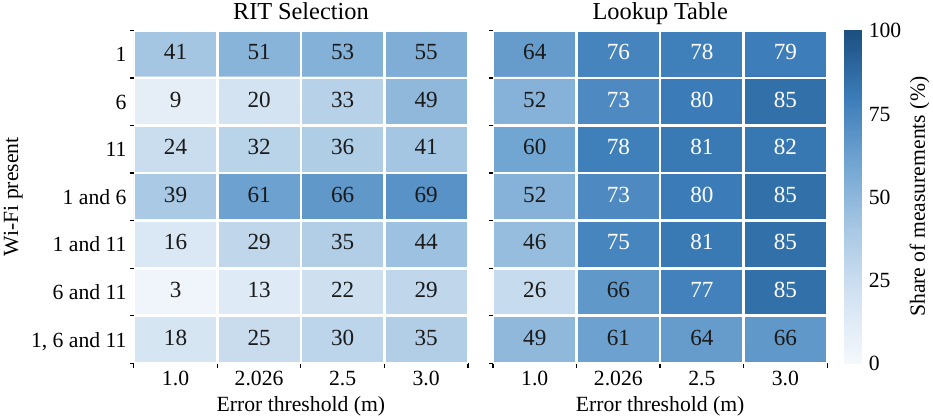}
\caption{Ranging success rates of RIT selection and the lookup table under seven Wi-Fi occupancy configurations. Rows indicate occupied Wi-Fi channels, columns indicate absolute-error thresholds, and cell values and colors represent the percentage of estimates below each threshold.
\vspace{-2mm}
}
\label{fig:grid}
\end{figure}

Fig.~\ref{fig:grid} compares ranging success rates for RIT selection and our proposed method across seven Wi-Fi occupancy configurations. The proposed method achieves higher success rates across all seven configurations. At the 2.026~m threshold, its average success rate is 71.4\%, compared with 33.4\% for RIT selection, an improvement of 38.0 percentage points. 
Note that the improvement persists across different error thresholds. 
Thus, the improvement is not confined to a specific resolution-cell metric.


\vspace{-1mm}
\subsection{Cost of Tone Selection}
The seven tone maps generated by Alg.~\ref{alg:rule}, shown in Table~\ref{tab:maps}, can be precomputed and stored for runtime lookup based on Wi-Fi occupancy, eliminating the need for online tone-selection optimization. At runtime, the device identifies the configuration and retrieves the corresponding map, avoiding an online search over candidate tone sets.


\begin{table}[t]
\caption{Lookup Table For Wi-Fi Channels 1, 6, 11}
\label{tab:maps}
\centering
\setlength{\tabcolsep}{4pt}
\footnotesize
\begin{tabular}{ll}
\toprule
Wi-Fi present & Channel tones \\
\midrule
1        & 22 26 30 34 37 41 45 49 53 57 61 64 68 72 76 \\
6        & 47 49 51 53 55 57 59 61 64 66 68 70 72 74 76 \\
11       & 2 5 9 12 15 18 22 26 28 32 35 38 41 45 48 \\
1, 6     & 47 49 51 53 55 57 59 61 64 66 68 70 72 74 76 \\
1, 11    & 22 26 27 28 29 31 33 35 37 39 41 42 44 46 48 \\
6, 11    & 2 3 5 6 8 9 11 12 13 15 16 18 19 21 22 \\
1, 6, 11 & 22 26 30 34 37 41 47 48 53 57 61 64 68 72 76 \\
\bottomrule
\end{tabular}
\end{table}

\vspace{-2mm}
\section{Experimental Validation}
\label{sec:public}
\vspace{-1mm}
To validate the simulation results, we use a public dataset containing I/Q measurements from commercial Bluetooth CS hardware at 60 positions with known distances \cite{wieme}. We evaluate 13 indoor positions spanning 3--22~m under line-of-sight (LOS) and non-line-of-sight (NLOS) conditions, excluding the 1 and 2~m positions. Table~\ref{tab:measured} compares our proposed method with RIT selection across six Wi-Fi occupancy configurations. Our proposed method achieves an average success rate of 52.0\%, compared with 23.3\% for RIT selection. The full-occupancy case is excluded because only eight nominally free tones remain, requiring both methods to select tones from occupied frequencies whose actual interference is not characterized in the public dataset. We evaluate this case separately using commercial hardware under Wi-Fi interference.  


\begin{table}[t]
\caption{Ranging Success Rate On The Public Bluetooth Channel Sounding Dataset}
\label{tab:measured}
\centering
\setlength{\tabcolsep}{5pt}
\footnotesize
\begin{tabular}{lccc}
\toprule
Wi-Fi Present    & RIT  & Lookup Table & $\Delta$ \\
\midrule
1      & 22.1\,\% & 48.3\,\% & $+26.2$ \\
6      & 12.5\,\% & 53.5\,\% & $+41.0$ \\
11     & 20.5\,\% & 63.9\,\% & $+43.4$ \\
1, 6   & 35.9\,\% & 53.5\,\% & $+17.6$ \\
1, 11  & 28.5\,\% & 32.1\,\% & $+3.6$  \\
6, 11  & 20.3\,\% & 60.6\,\% & $+40.3$ \\
\midrule
Mean & 23.3\,\% & 52.0\,\% & $+28.7$ \\
\bottomrule
\end{tabular}
\\[2pt]
\footnotesize Share within one resolution cell over 1{,}300 procedures per entry, from 13 indoor positions spanning 3 to 22~m in LOS and NLOS placements. 
\end{table}



\vspace{-2mm}
\subsection{Experimental Validation Using Commercial Hardware}
\label{sec:bench}
\begin{figure}[ht]
\centering
\includegraphics[width=0.45\textwidth]{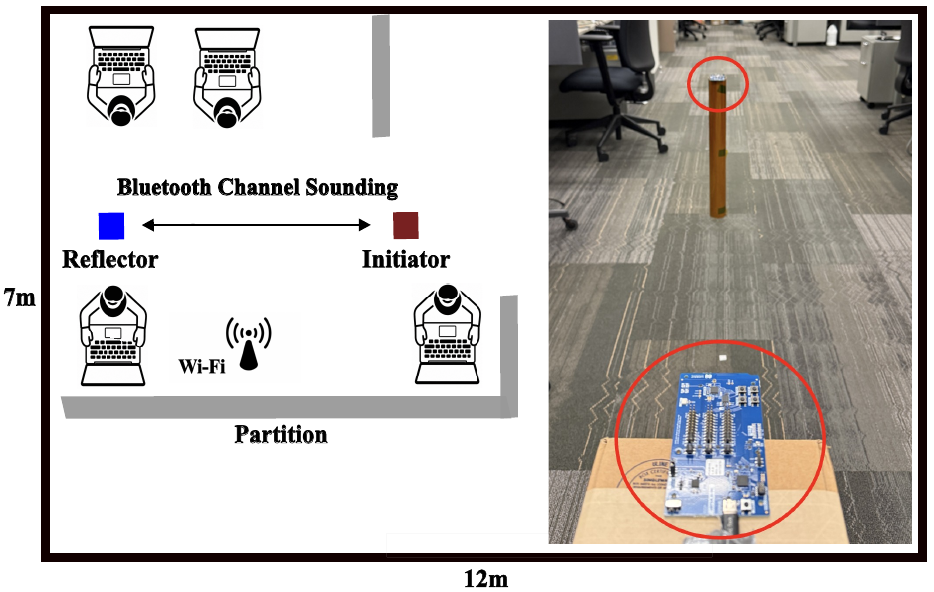}
\caption{Indoor Bluetooth CS measurement setup: floor plan of the 12 $\times$ 7~m office environment and photograph of the Nordic nRF54L15 DK development boards used as initiator and reflector. Wi-Fi channels 1, 6, and 11 are active during the measurements.
\vspace{-1mm}
}
\label{fig:testbed}
\end{figure}
We evaluate the full Wi-Fi occupancy case using two commercial Nordic nRF54L15 development kits in the 12 $\times$ 7~m office environment shown in Fig.~\ref{fig:testbed}, with Wi-Fi channels 1, 6, and 11 active throughout. We collect 17 captures at six separations between 4 and 10~m, evaluating the proposed map and three RIT maps with approximately 100 ranging procedures per map at each stationary placement.

\begin{figure}[t]
\centering
\includegraphics[width=0.85\columnwidth]{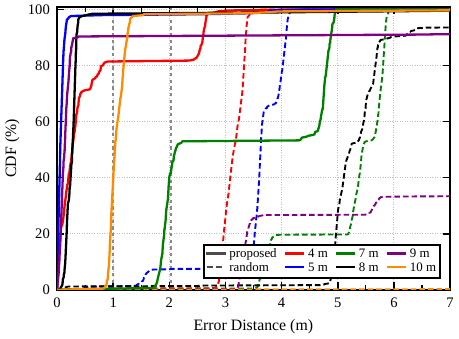}
\caption{Cumulative distributions of absolute ranging error measured using Nordic nRF54L15 DK hardware at six distances (4--10~m), with Wi-Fi channels 1, 6, and 11 active. Solid and dashed curves represent the proposed and RIT selection methods, respectively. Higher cumulative percentages at a given error threshold indicate better ranging accuracy.
\vspace{-1mm}
}
\label{fig:cdf}
\end{figure}

Fig.~\ref{fig:cdf} shows the cumulative distributions of absolute ranging error, where solid and dashed curves represent the proposed method and RIT selection, respectively. The proposed method achieves higher cumulative percentages across the tested distances. Overall, 83.9\% of its estimates fall within one resolution cell, compared with 1.6\% for RIT selection. The median absolute error decreases from 5.23 to 0.30~m.



\vspace{-1mm}
\section{Conclusion}
\vspace{-1mm}
Our paper presents an interference-aware tone-selection method for Bluetooth Channel Sounding. Our method uses occupied tones to construct a tone map in three steps: it identifies free tones, selects a suitable frequency window, and places the requested tones across that window. The resulting maps can be precomputed and stored for runtime lookup, avoiding online optimization and the need for ground-truth distance. Across indoor simulations, measurements from a public dataset, and using commercial hardware - our method shows substantial performance gains compared to simple interference avoiding benchmarks.


Future work will extend the experimental validation using commercial Bluetooth CS hardware across diverse indoor and outdoor environments, considering varying Wi-Fi occupancy and interference from other 2.4~GHz wireless technologies.

\end{document}